\documentclass[
  reprint,
  amsmath,amssymb,
  aps,
  prab,
  superscriptaddress,
  longbibliography
]{revtex4-2}
\usepackage{graphicx}
\usepackage{dcolumn}
\usepackage{bm}
\usepackage[caption=false]{subfig}
\usepackage{siunitx}
\usepackage[colorlinks=true,
            linkcolor=blue,
            citecolor=blue,
            urlcolor=blue]{hyperref}

\begin{document}

\title{Thermomechanical rf breakdown from magnetically focused field emission in high-gradient normal-conducting cavities}

\author{Dillon~C.~Merenich}
\affiliation{Northern Illinois University, DeKalb, Illinois 60115, USA}

\author{Tianhuan~Luo}
\email{tluo@lbl.gov}
\affiliation{Lawrence Berkeley National Laboratory, Berkeley, California 94720, USA}

\author{Xueying~Lu}
\email{xylu@niu.edu}

\affiliation{Northern Illinois University, DeKalb, Illinois 60115, USA}
\affiliation{Argonne National Laboratory, Lemont, Illinois 60439, USA}

\date{\today}

\begin{abstract}
Normal-conducting radiofrequency (rf) cavities for muon-collider ionization cooling must operate at high accelerating gradients in strong solenoidal magnetic fields, where rf breakdown can be enhanced by the magnetic focusing of field-emitted electrons. In this work, field-emitted electrons were tracked in the realistic field maps of rectilinear cooling-lattice cavities to test the validity of the previously developed localized-bombardment picture with simplified field maps. Under comparable reduced-field assumptions, the tracking results show good agreement with previous results. The full rf eigenmode fields and nonuniform solenoidal fields modify the idealized beamlet structure, producing rf phase dependent centroid shifts and broadened impact distributions from solenoid fringe fields. Nevertheless, the original model remains a useful framework for estimating limits on operating gradients. The thermal response is evaluated analytically using this model, with material properties varied to assess the coupled effects of heat transport and thermomechanical damage threshold. These results can inform cavity testing in strong solenoidal fields, muon ionization cooling channel designs, and other applications requiring high-gradient rf operation in magnetic fields.

\end{abstract}

\maketitle
\section{Introduction}
\label{sec:intro}

Muon colliders are a proposed path toward compact multi-TeV lepton colliders while avoiding the synchrotron-radiation limitations of circular electron--positron colliders \cite{mumuc,muoncollider_accelrd}. Because muons are produced as secondary particles with large initial phase-space volume and finite lifetime, rapid beam cooling is required before acceleration and collision. Ionization cooling provides a mechanism for reducing the beam phase space by passing the muon beam through absorbers while restoring longitudinal momentum with radiofrequency (rf) acceleration \cite{ic1,ic2}. A rectilinear six-dimensional cooling channel implements this concept using staged absorber, rf-cavity, and solenoidal-magnet configurations \cite{rectCool,IMCC-rect}.

The normal-conducting rf cavities in these channels must operate at high accelerating gradients inside strong solenoidal magnetic fields. This combination is challenging because magnetic fields can focus field-emitted electrons, producing localized pulsed heating and thermal stress \cite{bd-palmer,bd-diktys}. High-power cavity tests in multi-tesla solenoidal fields have shown reduced operating gradients and paired damage sites on opposing cavity walls, consistent with a mechanism where dark-current beamlets are guided by the magnetic field and deposit energy locally \cite{bd-bowring}. Earlier analytical models captured this process using simplified longitudinal rf and solenoidal fields, leading to a magnetic-field-dependent impact radius for the focused electron beamlet \cite{bd-diktys}. 

The influence of externally applied magnetic fields on vacuum and rf breakdown remains an active area of experimental and computational study. Measurements in dc vacuum gaps have shown that applied magnetic fields can modify dark-current behavior and reduce the observed breakdown voltage \cite{lebedynskyi-Bbreakdown}. More recent particle-in-cell and finite-element simulations have examined the transport of field-emitted electrons in magnetic fields up to several tens of tesla, demonstrating that the magnitude and orientation of the applied field can strongly influence electron focusing and the resulting localized heating of the opposing surface \cite{koitermaa-Bheating}. Complementary rf studies are developing dedicated cavity test facilities for operation in multi-tesla magnetic fields and investigating the physical mechanisms that determine breakdown under combined rf electric and static magnetic fields \cite{imcc-rf-test,imcc-breakdown-studies, grudiev-aac26, snively-usmcc25}. Recent hydrodynamic modeling has further examined the deformation of field-emitting surface features under combined rf and dc magnetic fields, extending these studies toward the evolution of the emitting surface itself \cite{plouin-mhd}. These efforts emphasize the need to understand the roles of field emission, electron transport, surface heating, and emitter evolution in determining high-field performance in the presence of an external magnetic field.

Recent studies of copper and copper alloys indicate that breakdown performance depends on more than a single material property. CuAg and cryogenic copper have been investigated because alloying and reduced-temperature operation modify their electromagnetic, thermal, and mechanical properties \cite{cuag-cband,cuag-cryo}. Additionally, Density Functional Theory (DFT) studies of solid-solution-strengthened copper have connected alloy-induced changes in lattice and elastic properties to strengthening metrics such as critical resolved shear stress, while also predicting changes in thermal transport \cite{alloy1,alloy2}. More generally, solute atoms can impede dislocation motion and increase the stress required for slip \cite{sss1,slip-disloc}, while changes in composition can simultaneously modify electron and phonon transport. Mesoscale modeling has further shown that electric fields and elevated temperatures can act together to promote surface instability and the formation of sharp breakdown precursors, with thermoelastic stresses reducing the critical field required for this instability \cite{thermal-precursor}. Experimental studies likewise indicate that high-rf field conditioning involves changes in the near-surface and subsurface defect structure and cannot necessarily be described by surface hardening alone \cite{ccfe,subsurf}. These results are beginning to support a coupled physical picture in which field emission and surface level effects lead to arcing that triggers breakdown, while thermal transport, dislocation activity, and the evolving material structure determine precursors and responses to external loadings. The available data do not yet establish a general relationship between hardening and breakdown resistance, but they motivate treating conductivity, hardness, or thermal diffusivity as interconnected rather than independent predictors.

This work applies the localized-bombardment framework to realistic rf cavities in the rectilinear muon-collider cooling lattice and evaluates the thermomechanical response to the resulting focused electron bombardment. Field-emitted electrons are tracked using full rf electric- and magnetic-field maps and spatially varying external solenoidal magnetic-field maps to test whether the simplified model based on a longitudinal rf electric field and uniform longitudinal external magnetic field \cite{bd-diktys} remains relevant when transverse rf components and nonuniform lattice fields are included. The resulting localized thermal-loading is then studied with a previously developed analytical temperature-rise model, where field emission and magnetic focusing define the heat-source size and intensity, and the material properties determine heat storage, thermal diffusion, and thermal-stress response. Copper-like material-property variations, motivated by NIST copper data and DFT-based copper-alloy studies, are used to examine how heat transport, rf pulse length, field enhancement, and thermomechanical threshold affect the predicted gradient limit. The goal is not to predict an absolute breakdown limit for a particular processed surface but to provide a physics-based context for interpreting high-field rf breakdown trends and comparing material and operating-condition choices. These results are useful for muon-collider ionization cooling cavities as well as other normal-conducting cavities operating in magnetic fields.

The remainder of this paper examines the effects of realistic electromagnetic field components on dark-electron trajectories and investigates how material properties influence resilience to breakdown in this electron-bombardment regime. The paper is organized as follows. Section~\ref{sec:theory} reviews the analytical framework for magnetically focused field emission, localized energy deposition, and the resulting thermomechanical response. Section~\ref{sec:tracking} then examines the magnetic-focusing assumptions of this framework by tracking field-emitted electrons using all electric and magnetic components of the rf eigenmode together with the longitudinal and transverse components of the external magnetic field. Section~\ref{sec:thermal_response} applies the analytical framework using the modeling assumptions developed for the present study to evaluate the thermal response and its dependence on material and operating parameters. The limitations of the model and broader implications of the results are then discussed in Sec.~\ref{sec:discussion}, and the conclusions are presented in Sec.~\ref{sec:conclusion}.

\section{Theory of Magnetically Focused Field Emission and Coupled Thermomechanical Response}
\label{sec:theory}

The localized-bombardment model of Refs.~\cite{bd-palmer,bd-diktys} provides a framework for relating field emission in an rf cavity to localized thermal loading in the presence of an external magnetic field. In this picture, electrons emitted from locally enhanced electric-field regions are accelerated by the rf electric field and transported across the cavity. A sufficiently strong longitudinal external magnetic field confines the emitted charge into localized beamlets, increasing the deposited power density at the impact surface. The resulting temperature rise can generate thermomechanical stresses that exceed the elastic limit of the material and lead to cyclic plastic deformation. The central elements of this framework are summarized here before examining its applicability to the full rf eigenmode fields and external lattice magnetic field of the rectilinear cooling channel.

At an emitting surface, the local rf electric field, $E_{\mathrm{rf}}(\varphi)=E_0\cos\varphi$,
where $E_0$ is the local surface-field amplitude and $\varphi$ is the rf phase. Electron emission from a locally enhanced surface field is described by the Fowler--Nordheim relation,
\begin{equation}
j_{em}(\varphi) = 1.54 \times 10^{-6} \frac{(\beta_{FE}E_{\mathrm{rf}})^{2}}{\phi}
\exp\left(\frac{-6.53\times10^9 \phi^{1.5}}{\beta_{FE}E_{\mathrm{rf}}}\right),
\label{fn_eq}
\end{equation}
where $\beta_{\mathrm{FE}}$ is the local field-enhancement factor and $\phi$ is the material work function. The corresponding emission current is determined by integrating the current density over the emitting area. Because of the strong dependence of Fowler--Nordheim emission on the enhanced surface field, emission is concentrated near regions of large local field enhancement.

In the reduced magnetic-focusing model of Ref.~\cite{bd-diktys}, the emitted electrons were transported through an rf cavity containing a longitudinal rf electric field and a spatially uniform longitudinal external magnetic field. Close to the emitting asperity, transverse momentum arises from the local enhanced fields and space-charge forces. As the electrons accelerate away from the source, the space-charge contribution weakens and the external magnetic field confines the beamlet during transport. The resulting rms collision radius was parameterized as
\begin{equation}
R_f=\frac{22.6\overline{I}_{\mathrm{em}}^{1/3}}{B\mathrm{{ext}}},
\label{eq:beamlet_radius}
\end{equation}
where $R_f$ is in $\mu$m, the rf-cycle-averaged emission current $\overline{I}_{\mathrm{em}}$ is in $\mu$A, and the external magnetic field $B\mathrm{{ext}}$ is in tesla. This relation reflects the competition between current-dependent transverse beamlet expansion and magnetic confinement and retains the characteristic inverse dependence of the impact radius on the external magnetic field. The relation was obtained for an 805-MHz pillbox cavity with a spatially uniform longitudinal solenoidal field.

When the focused electrons strike the cavity surface, their kinetic energy is deposited as they penetrate the material. The volumetric deposited power can be expressed generally as
\begin{equation}
W(z)=
\frac{I_{\mathrm{em}}}{q\pi R_f^2}
\frac{dE}{dz},
\label{w}
\end{equation}
where $q$ is the magnitude of the electron charge and $dE/dz$ is the collisional energy loss per unit depth. Thus, the deposited-power density depends both on the energy deposited by the incident electrons and on the transverse area over which the beamlet is focused. Because $R_f$ decreases with increasing magnetic field, magnetic focusing can increase the local deposited-power density without directly changing the incident electron current or energy.

The resulting thermal response is described using the heat-conduction equation in cylindrical coordinates for a localized source. The thermal diffusivity is
\begin{equation}
\alpha_d=\frac{\kappa}{\rho C_p},
\label{thermal_diffusivity}
\end{equation}
where $\kappa$ is the thermal conductivity, $\rho$ is the material density, and $C_p$ is the specific heat capacity. For an axisymmetric source in a semi-infinite material, the heat equation can be reduced using radial and longitudinal Green's functions \cite{bd-diktys,greenfunc}. The radial Green's function is
\begin{equation}
G_r(r,r';t,t')=
\frac{
\exp\left[-\frac{r^2+r'^2}{4\alpha_d(t-t')}\right]
}{
4\pi\alpha_d(t-t')
}
I_0\left(
\frac{rr'}{2\alpha_d(t-t')}
\right),
\label{eq:radial_green}
\end{equation}
and the longitudinal Green's function is
\begin{equation}
G_z(z,z';t,t')=
\frac{
\exp\left[-\frac{z^2+z'^2}{4\alpha_d(t-t')}\right]
}{
\sqrt{\pi\alpha_d(t-t')}
}
\cosh\left(
\frac{zz'}{2\alpha_d(t-t')}
\right),
\label{eq:longitudinal_green}
\end{equation}
where $I_0$ is the modified Bessel function of the first kind.

The resulting temperature rise is
\begin{equation}
\begin{aligned}
\Delta T(r,z,t;E_0,B)
&=
\frac{2\pi\alpha_d}{\kappa}\times \\
&
\int_0^t
\int_0^D
\int_0^{R_f}
G_rG_zWr'dr'dz'dt',
\end{aligned}
\label{trise}
\end{equation}

where $R_f$ defines the transverse extent of the focused heat source, $D$ is the characteristic penetration depth, and $W$ is the volumetric deposited-power density. The characteristic thermal diffusion length is
\begin{equation}
\ell_d\sim\sqrt{\alpha_d\tau_{\mathrm{rf}}},
\label{diffusion_length}
\end{equation}
so the resulting temperature rise depends on the spatial concentration of the deposited energy and the distance over which heat diffuses during the rf pulse.

The calculated temperature rise can be compared with a thermomechanical threshold based on the onset of constrained cyclic plastic deformation \cite{pulsed-heating-rf,musal},
\begin{equation}
\Delta T_s\approx
2\frac{(1-\nu)\sigma_y}
{E_{\mathrm{mod}}\alpha_{\mathrm{th}}},
\label{tthresh}
\end{equation}
where $\nu$ is Poisson's ratio, $\sigma_y$ is the yield strength, $E_{\mathrm{mod}}$ is Young's modulus, and $\alpha_{\mathrm{th}}$ is the thermal expansion coefficient. The factor of two approximates loading through both compression and tension during a thermal cycle. The threshold is used as a comparative criterion for the onset of cyclic plastic deformation, not as a complete microscopic criterion for rf breakdown.

The central assumption of this framework is that the external magnetic field confines the emitted electrons into a localized impact region whose characteristic size can be represented by Eq.~\eqref{eq:beamlet_radius}. That relation was obtained using a simplified longitudinal rf electric field and a static, spatially uniform longitudinal external magnetic field. In the following section, dark electrons are tracked using the full rf electric- and magnetic-field maps together with the spatially varying external magnetic-field maps to determine how more complex electromagnetic field histories modify their transport and collision distributions.

\section{Tracking Simulations of Field-Emitted Electrons in Cavities in Muon Rectilinear Cooling Channels}\label{sec:tracking}
\label{sec:sims}

\begin{figure}[b!]
\centering
\includegraphics[width=0.95\linewidth]{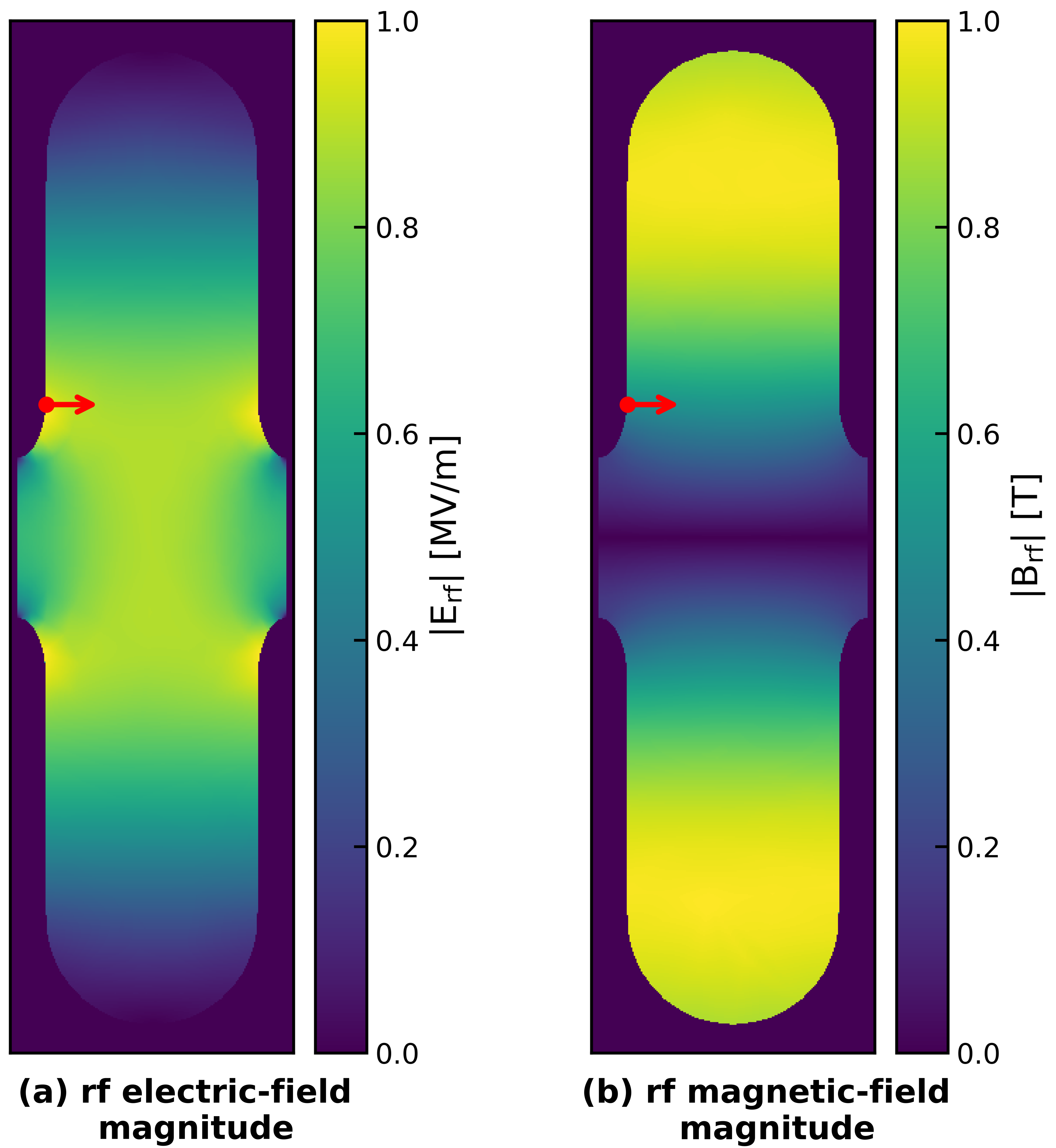}
\caption{
Normalized rf eigenmode fields for the Stage~B8 cavity from the cooling-channel design of Refs.~\cite{IMCC-rect,barbagallo-cavity}, calculated using the CST Eigenmode Solver.
Panel (a) shows the rf electric-field magnitude, and panel (b) shows the rf magnetic-field magnitude. The dot indicates the selected field-emission location at the peak surface electric field, and the arrow indicates the local surface-normal direction used to initialize the emitted electrons.
}
\label{fig:rfmaps}
\end{figure}

The analytical framework introduced in Sec.~\ref{sec:theory} describes the magnetic focusing of field-emitted electrons using simplified longitudinal rf and external magnetic fields. The tracking simulations presented here examine how the electron transport and collision distributions are modified when the simplified fields are replaced by the full rf cavity fields and the spatially varying external magnetic fields of the rectilinear cooling channel.

The rf cavities used throughout the rectilinear cooling channel follow the 704-MHz cavity design described in Ref.~\cite{barbagallo-cavity}. The cavity is an elliptical, pillbox-like structure operating in a $\mathrm{TM}_{010}$-like mode at 704~MHz. Each cavity is enclosed with thin beryllium windows, which provide resilience to breakdown while permitting a higher on-axis field for a given supplied power and a higher shunt impedance \cite{bd-bowring,Derun-Be,diktys-lattice,barbagallo-cavity}.

While the cavity design is common across the cooling channel, the rectilinear lattice is divided into a sequence of stages with different operating and focusing parameters as the muon-beam emittance is reduced. The external magnetic-field configuration and required accelerating gradient therefore vary along the channel \cite{IMCC-rect}. Stage~B8 is selected for the present tracking study because it combines a relatively high accelerating gradient with a compact cavity aperture, producing large surface electric fields near the cavity iris. For Stage~B8, the nominal accelerating gradient is 27.71~MV/m, and the maximum rf electric field near the iris is approximately 30~MV/m. The normalized rf electric- and magnetic-field magnitudes used in the tracking calculation are shown in Fig.~\ref{fig:rfmaps}. The rf electric- and magnetic-field maps are obtained from the cavity design of Ref.~\cite{barbagallo-cavity}, while the external solenoidal field maps introduced below are taken from the Stage~B8 rectilinear cooling-lattice design of Ref.~\cite{IMCC-rect}. The emission source is placed at the location of the peak rf electric field on the cavity surface, with the local emission direction normal to the surface.

\begin{figure}[tb!]
\centering
\includegraphics[width=\linewidth]{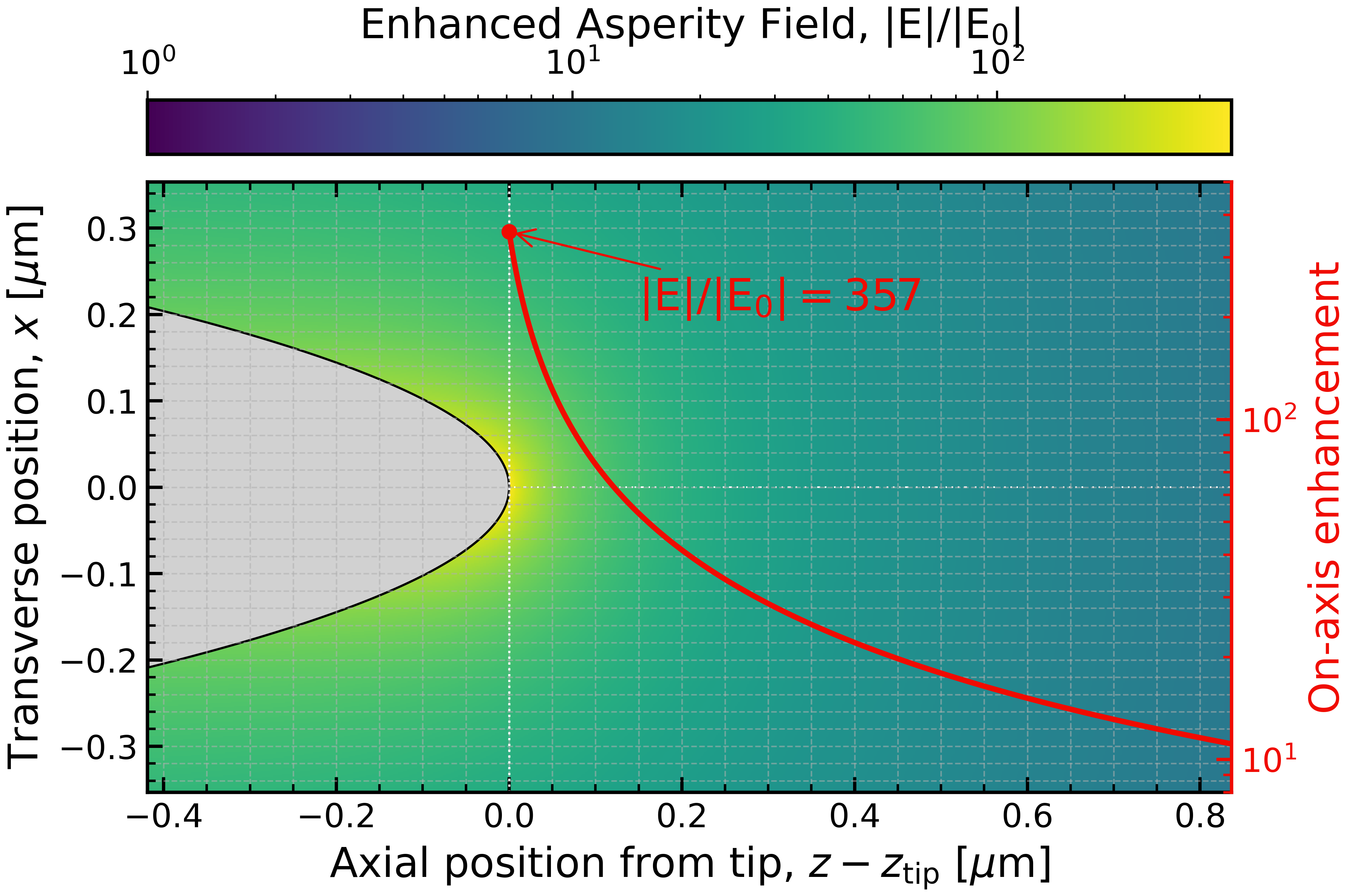}
\caption{
Field-enhancement distribution near the tip of the analytical prolate-spheroid asperity used to initialize the field-emission source. The overlaid curve (right axis) shows the electric field enhancement along the asperity major axis.
}
\label{fig:asperity-enhancement}
\end{figure}

The local field-emission source is initialized as a single emitter using an analytical prolate-spheroid asperity model consistent with the field-enhancement treatment of Ref.~\cite{bd-diktys}. The emitting feature has a semi-major axis $a=60~\mu\mathrm{m}$ and a semi-minor axis $b=1.77~\mu\mathrm{m}$. The imported rf electric field at the selected emission site provides the macroscopic rf electric-field amplitude and direction for the analytical asperity calculation. The local field-enhancement factor is defined as the ratio of the enhanced electric field magnitude to the macroscopic rf electric field magnitude, or $\mathrm{|E|/|E_0|}$. For the geometry used here, the calculated apex enhancement is $\beta_{\mathrm{FE}}=357$. The corresponding near-tip field enhancement distribution is shown in Fig.~\ref{fig:asperity-enhancement}. The instantaneous emission current density is evaluated using the Fowler--Nordheim relation in Eq.~\eqref{fn_eq} during the accelerating phase of the rf cycle. The emitted charge is discretized in time and space to form a time-dependent bunch of 500,000 macroparticles. Each particle is initialized with a kinetic energy of 1~eV directed along the local surface normal direction. The detailed field-enhancement calculation, temporal sampling distribution, and macroparticle allocation procedure are provided in Appendix~\ref{app:field_enhancement}.

\begin{figure}[tb!]
\centering
\includegraphics[width=\linewidth]{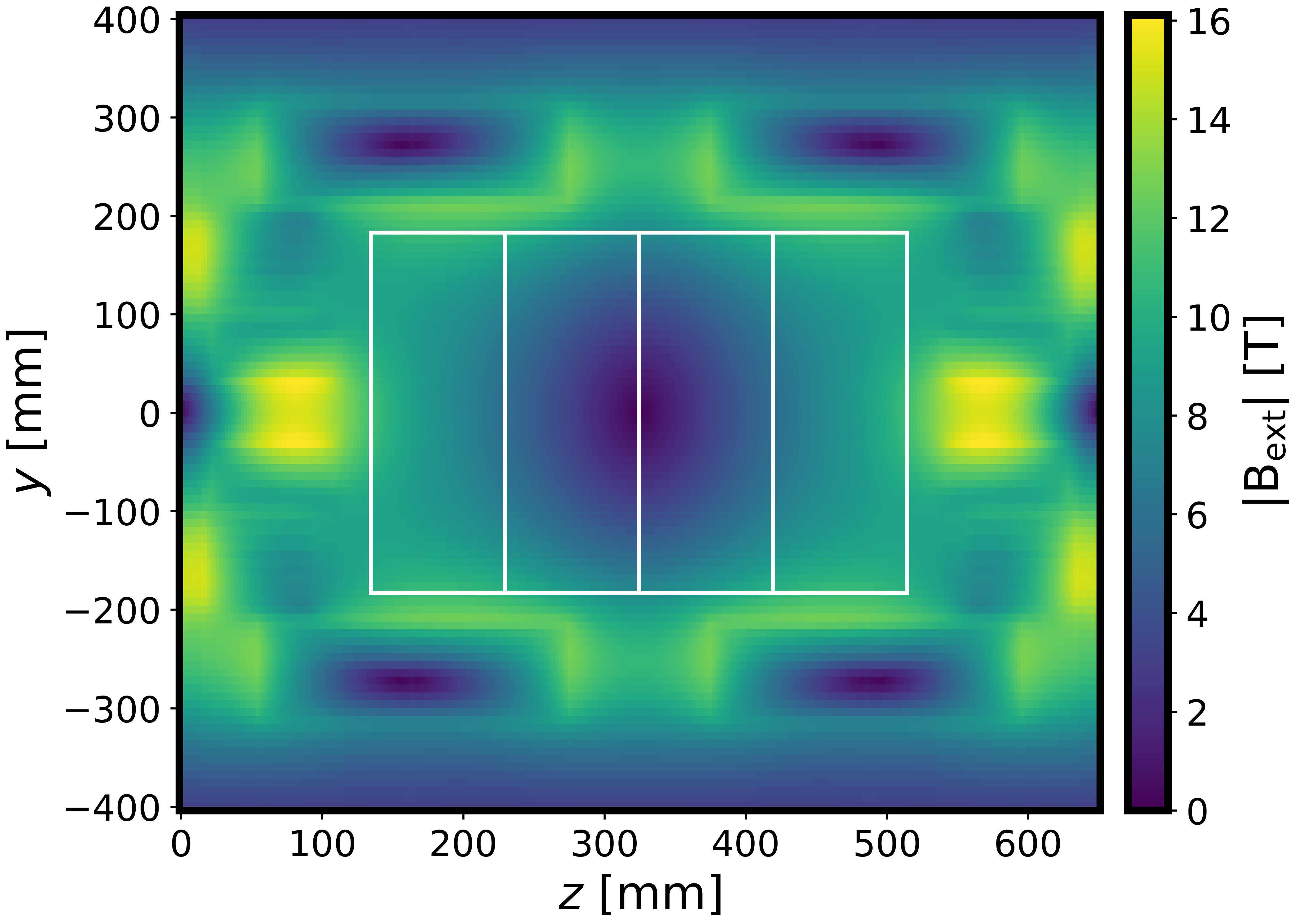}
\caption{Magnitude of the Stage~B8 lattice external magnetic-field map, produced using the design data from Ref.~\cite{IMCC-rect}. The white boxes indicate the cavity positions within the cooling cell.
}
\label{fig:solenoidB8}
\end{figure}

The external solenoid magnetic field is taken from the Stage~B8 cooling-lattice design of Ref.~\cite{IMCC-rect}. The external magnetic field is designed to meet the emittance budget required to achieve the cooling goals at each stage. Figure~\ref{fig:solenoidB8} shows the magnitude of the external magnetic-field map for a Stage~B8 cooling cell in the plane $x=0$, with the cavity positions indicated by white boxes. The coordinates $x$, $y$, and $z$ represent the horizontal, vertical and longitudinal directions, respectively, with $z$ being the direction of forward travel by the muon beam. The solenoidal field is not uniform across the cavity region: the longitudinal solenoidal component varies throughout each cavity cell, and transverse components are present. Across the four cells, the longitudinal field is symmetric and mirrored about the center of the lattice section, with the two cavities on one side experiencing positive $B_z$ and the two cavities on the opposite side experiencing the mirrored negative-polarity field. For the present field-emission tracking study, we consider the first cavity cell, which corresponds to the largest magnetic fields experienced within the cavity. The remaining cavities are expected to exhibit the same basic localized-bombardment behavior, with the dark electron dynamics scaling with the field magnitudes in the subsequent cells.

The particle tracking simulation is divided into two phases to account for the strong, spatially localized electric-field enhancement near the prolate-spheroidal emitter. Space-charge forces are included in both phases. In Phase I, the cavity rf electric field is combined with the local field enhancement calculated from the emitter geometry shown in Fig.~\ref{fig:asperity-enhancement}. This phase captures the initial acceleration and transverse expansion of the high-charge-density electron beamlet within a localized region near the emitter. The resulting particle distribution is then transferred to Phase II, in which the localized asperity electric-field contribution is excluded and the particles are propagated through the cavity-scale electromagnetic fields. The rf electric- and magnetic-field components and external solenoidal magnetic-field components included in Phase II are varied among the three configurations described below. Separating the simulation into these two phases allows the mesh size and time step to be controlled independently in each phase, thus reducing the computational cost while improving numerical accuracy.

\begin{figure}[tb!]
\centering
\includegraphics[width=\linewidth]{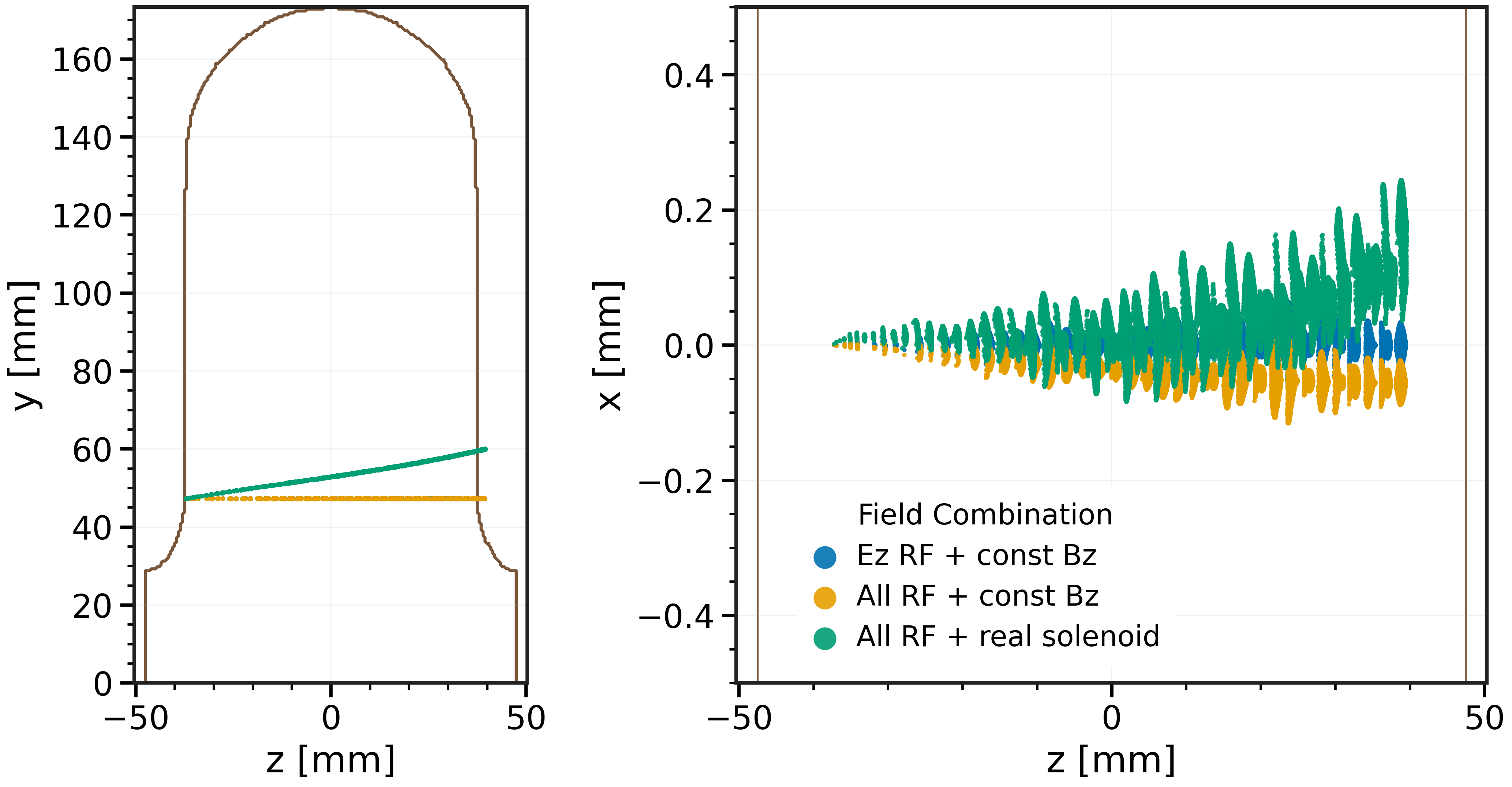}
\caption{
Beamlet trajectory snapshots shown in the $z$--$y$ and $z$--$x$ projections, using cavity coordinates with the cavity center defined as the origin. The blue markers correspond to particles tracked with only $E_{z,\mathrm{rf}}$ and uniform $B_{\mathrm{ext},z}$. The orange markers correspond to particles tracked with the full rf eigenmode fields and uniform $B_{\mathrm{ext},z}$. The green markers correspond to particles tracked with the full rf eigenmode fields and the spatially varying external Stage~B8 lattice field. The cavity boundaries are indicated by black lines.
}
\label{fig:trajectory-overview}
\end{figure}

\begin{figure*}[tb!]
\centering
\includegraphics[width=\textwidth]{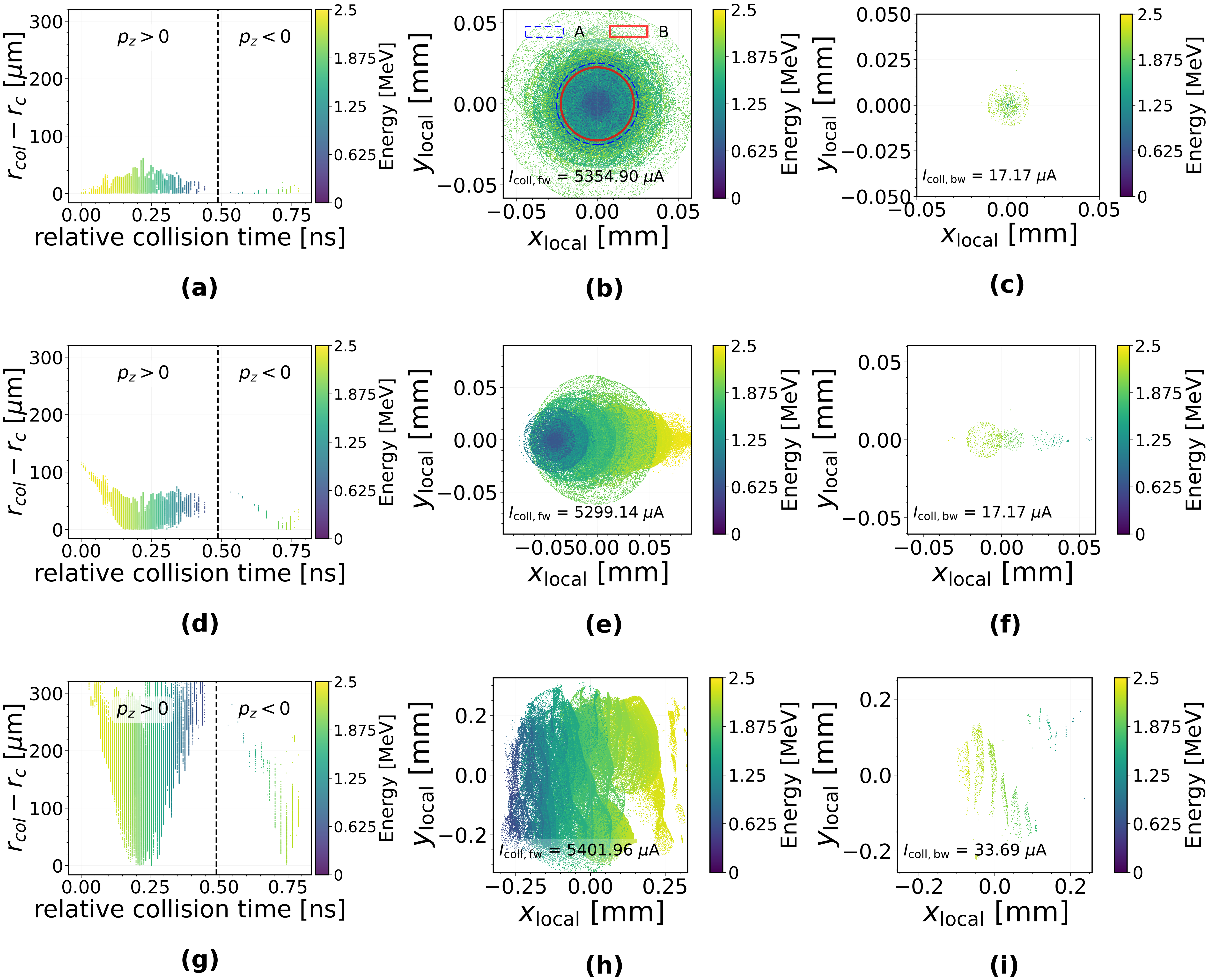}
\caption{
Comparison of the simulated collision profiles for the three field configurations. The top row, panels (a)--(c), corresponds to the reduced-field configuration with only $E_{z,\mathrm{rf}}$ and a constant longitudinal external magnetic field. In panel (b) the dashed circle labeled A shows the collision radius predicted by Eq.~\eqref{eq:beamlet_radius}, while the solid circle labeled B shows the rms radius obtained from the tracked collision distribution. The middle row, panels (d)--(f), includes the full rf eigenmode fields with the same constant external magnetic field. The bottom row, panels (g)--(i), includes the full rf eigenmode fields and the spatially varying Stage~B8 lattice magnetic field. The left column shows collision time and local collision radius, with color indicating collision energy. The center and right columns show the forward- and backward-propagating collision distributions, respectively. The local surface coordinates $x_{\mathrm{local}}$ and $y_{\mathrm{local}}$ are defined relative to the charge-weighted mean collision position for each propagation direction. The legends on the $x_{\mathrm{local}}$ and $y_{\mathrm{local}}$ figures indicated to corresponding forward and backward current, $I_{coll,fw}$ and $I_{coll,bw}$, respectively. 
}
\label{fig:collision-comparison}
\end{figure*}

Three field configurations are considered in the tracking simulations. The first tracking case uses only the longitudinal rf electric field, $E_{z,\mathrm{rf}}$, and a static, spatially uniform longitudinal external magnetic field, $B_{\mathrm{ext},z}=15.8$~T. This configuration reproduces the field structure assumed in the simplified magnetic-focusing model in Ref.~\cite{bd-diktys}. The resulting trajectories remain close to the idealized beamlet picture, with particles transported to neighboring cavity surfaces in compact localized groups. The collision time profile in Fig.~\ref{fig:collision-comparison}(a) separates the forward and backward populations by arrival time and local radius. It also shows that the impact energy is primarily correlated with rf emission phase, while the collision radius varies with the phase-dependent current and associated space-charge expansion.

The forward- and backward-propagating currents reported in Fig.~\ref{fig:collision-comparison}(b) and Fig.~\ref{fig:collision-comparison}(c) are calculated from the total charge carried by each population to the cavity surface per rf cycle, $I_{\mathrm{coll}}=Q_{\mathrm{coll}}f_{\mathrm{rf}}$. For the forward-propagating population, the tracked rms collision radius is $22.6~\mu\mathrm{m}$, compared with $R_f=25.2~\mu\mathrm{m}$ from the analytical prediction according to Eq.~\eqref{eq:beamlet_radius}. The simulated collision current of 5.35 mA also agrees well with the analytical prediction of 5.44 mA in the forward direction. The backward-propagating population accounts for less than 1\% of the total collision current. Its contribution to the total deposited power is therefore small, even though individual backward-propagating electrons can reach higher impact energies. Nevertheless, these impacts demonstrate that electrons emitted from a single localized source can reach regions on both neighboring cavity surfaces along the magnetic-field direction. 

The reduced-field configuration therefore provides a reference for evaluating the effects of the additional rf and lattice-field components. The presence of both forward- and backward-propagating impacts shows that electrons emitted from a single localized source can bombard both the forward and backward surfaces during different portions of the rf cycle. Repeated localized bombardment at these positions could contribute to the formation of paired damage or breakdown sites on opposing cavity surfaces.

The second tracking case includes all electric and magnetic components of the rf eigenmode while retaining the static, spatially uniform longitudinal external magnetic field, $B_{\mathrm{ext},z}=15.8$~T. This configuration isolates the effects of the transverse rf components from those produced by spatial variation of the external magnetic field. Figure~\ref{fig:collision-comparison}(d) shows that the collision energy remains correlated with arrival time, while the radial structure includes both the finite beamlet width and phase-dependent motion of the collision centroid. The forward distribution in Fig.~\ref{fig:collision-comparison}(e) shows a correlated progression of the beamlet centroids along the local surface-tangent direction. For the selected emission site near the $x=0$ plane, this direction is set by the local azimuthal rf magnetic field and its associated transverse Lorentz force. The backward distribution in Fig.~\ref{fig:collision-comparison}(f) is modified by the same phase-dependent transverse motion, with the impacts remaining concentrated near the emission site. The collision current, impact energy, and local beamlet radius associated with each rf emission phase remain consistent with the reduced-field case. The primary effect of the full rf eigenmode fields is therefore a phase-dependent displacement of the beamlet centroids rather than a significant change in the properties of the individual beamlets.

The final tracking case retains all electric and magnetic components of the rf eigenmode and replaces the uniform longitudinal external magnetic field with the spatially varying Stage~B8 lattice field. Within the first cavity, the longitudinal component of the external magnetic field varies from approximately 11~T to 5~T, and transverse external magnetic-field components are also present. Relative to the full-rf, constant-solenoid case, Fig.~\ref{fig:collision-comparison}(g) shows a broader time-dependent radial structure. Because the local radius is measured from the charge-weighted center of the complete impact distribution, this quantity combines phase-dependent centroid shifts with changes in the widths and shapes of the individual beamlets. The collision energy remains primarily correlated with arrival time and the rf acceleration history, whereas the spatial impact pattern becomes more sensitive to the external magnetic-field history sampled during transport. The forward distribution in Fig.~\ref{fig:collision-comparison}(h) spans a wider range of centroid positions and no longer retains the approximately circular beamlet structure observed in the constant-solenoid cases. Across the first cavity, the decreasing axial external magnetic field weakens transverse confinement, while the strong transverse components of the external magnetic field in the lattice introduce additional radial and azimuthal deflections that deform the individual beamlets. The backward impacts in Fig.~\ref{fig:collision-comparison}(i) are modified by the same field structure but remain concentrated near the emission-side region. Thus, the impacts remain localized relative to the cavity dimensions, but their detailed spatial distribution can no longer be characterized as a collection of approximately circular beamlets with a single representative radius.

The tracking simulations provide the justification for using the analytical heating model in the next section. In the reduced-field configuration for which the original bombardment model was developed, the simulated forward rms collision radius agrees very well with the scaling relation in Eq.~\eqref{eq:beamlet_radius}. When the full rf eigenmode and spatially varying lattice fields are included, the impacts remain localized relative to the cavity dimensions, although their centroids shift and the realistic external magnetic field distorts the approximately circular beamlet structure. The analytical heating model will therefore be used as a controlled description of the characteristic size and concentration of the deposited power, rather than as a reconstruction of the full impact morphology.

\section{Modeling Thermal Response to Focused Electron Bombardment}
\label{sec:thermal_response}
Building on the analytical framework of Sec.~\ref{sec:theory} and the tracking results of Sec.~\ref{sec:tracking}, this section evaluates the thermal response to magnetically focused electron bombardment using the reduced model of Ref.~\cite{bd-diktys}. The tracking simulations show that the electron impacts remain localized relative to the cavity dimensions, although the full rf and lattice fields modify the detailed collision morphology. The analytical beamlet description is therefore retained as a characteristic representation of the localized deposited-power source. The modeling assumptions used to evaluate the collision energy, energy deposition, and temperature rise are described first, after which the thermomechanical criterion of Eq.~\eqref{tthresh} is used to examine how the predicted limiting electric field depends on material properties, field enhancement, magnetic field, and rf pulse length.
\subsection*{Model assumptions and implementation}
\label{sec:thermal_model}
The thermal-response calculation evaluates the framework introduced in Sec.~\ref{sec:theory} using a phase-dependent description of the field-emitted electron bombardment. The emission current is calculated from Eq.~\eqref{fn_eq} using an emitting area $A_{\mathrm{em}}=10^{-14}$~m$^2$. Unless otherwise stated, the thermal model uses a baseline field-enhancement factor $\beta_{\mathrm{FE}}=370$. The characteristic impact radius entering Eqs.~\eqref{w} and \eqref{trise} is calculated from Eq.~\eqref{eq:beamlet_radius}.

The collision energy, $E_{\mathrm{coll}}(\varphi)$, is calculated separately for each emission phase by integrating the one-dimensional electron motion across a uniform $95$-mm cavity gap under the applied longitudinal electric field $E_0\cos\left(\varphi+\omega_{\mathrm{rf}}t\right)$, where $\omega_{\mathrm{rf}}=2\pi f_{\mathrm{rf}}$ and $f_{\mathrm{rf}}=704$~MHz. Electrons are initialized at the emitting surface with a kinetic energy of $1$~eV. The field-enhancement factor is applied only in the field-emission calculation and is not applied to the macroscopic rf electric field used to accelerate electrons across the cavity gap. Emission phases for which the electron reverses direction or does not reach the opposing surface do not contribute to the forward deposited-power source.

For each impacting phase, the projected penetration depth is estimated using the empirical relation employed in Refs.~\cite{bd-palmer,bd-diktys},
\begin{equation}
D(\varphi)=0.0267\frac{A_{\mathrm{at}}}{\rho Z^{0.89}}
\left[E_{\mathrm{coll}}(\varphi)\right]^{1.67},
\label{eq:penetration_depth}
\end{equation}
where $D$ is in $\mu$m, $E_{\mathrm{coll}}$ is expressed in keV, $A_{\mathrm{at}}$ is the atomic mass in g/mol, $Z$ is the atomic number, and $\rho$ is the material density in g/cm$^3$. The collisional energy loss within the material is calculated using the Bethe--M{\o}ller expression \cite{bethe,berger-seltzer},
\begin{equation}
\frac{dE}{dz}=
\frac{2\pi N_A r_e^2m_ec^2\rho Z}
{A_{\mathrm{at}}\beta_e^2}
\left[
\ln\left(
\frac{E^2}{I_{\mathrm{exc}}^2}
\left(1+\frac{\tau}{2}\right)
\right)
+F^{-}(\tau)-\delta
\right],
\label{eq:bethe_moller}
\end{equation}
where $\tau=E/(m_ec^2)$, $\beta_e$ is the electron velocity normalized to $c$, and $I_{\mathrm{exc}}$ is the material mean excitation energy. The electron correction $F^{-}(\tau)$ follows the standard Bethe--M{\o}ller treatment \cite{bethe,berger-seltzer}, and the density-effect correction is set to $\delta=0$. The electron kinetic energy is reduced according to this stopping power until the energy is exhausted or the projected penetration depth is reached.

To evaluate the volumetric source in Eq.~\eqref{w}, the correlation between emission current, collision energy, and depth-dependent energy loss is retained for each rf phase before averaging over the rf cycle. The source used in the thermal calculation is therefore
\begin{equation}
W(z;E_0,B)=
\frac{1}{q\pi R_f^2}
\left\langle
I_{\mathrm{em}}(\varphi;E_0)
\frac{dE(\varphi,z)}{dz}
\right\rangle_{\mathrm{rf}},
\label{eq:w_rf_average}
\end{equation}
where $\langle\cdots\rangle_{\mathrm{rf}}$ denotes an average over one complete rf period. Phases that do not reach the opposing surface contribute zero deposited power. Consistent with the source geometry in Sec.~\ref{sec:theory}, the deposited power is taken to be uniform within the circular area $\pi R_f^2$ and zero outside the impact region.

The rf-cycle-averaged source is held constant during the pulse flattop. The baseline pulse length is $\tau_{\mathrm{rf}}=20~\mu\mathrm{s}$, with other pulse lengths considered separately. The calculation neglects the cavity filling and decay intervals and assumes that $E_0$, $\beta_{\mathrm{FE}}$, $A_{\mathrm{em}}$, and the material properties remain fixed during each pulse. The temperature rise is then evaluated using Eq.~\eqref{trise}, with the peak response taken at the center of the impact region and the material surface, $r=0$ and $z=0$.

The safe operating field, $G_{\mathrm{safe}}$, is defined as the local rf surface-field amplitude for which the calculated peak temperature rise reaches the thermomechanical threshold in Eq.~\eqref{tthresh},
\begin{equation}
\Delta T(r=0,z=0,t;G_{\mathrm{safe}},B)= \Delta T_s.
\label{eq:sog_condition}
\end{equation}
This criterion is used consistently in the following subsections when varying the material and operating parameters.
\subsection*{Breakdown gradient dependence on the thermal and mechanical properties}
\label{sec:relative_material_inputs}

To isolate how the thermal response depends on individual material inputs, the thermal conductivity, specific heat capacity, and density are varied individually. The material properties are selected relative to copper (Cu) property set at 300~K \cite{nist-data}, while all other model inputs are held fixed. Thus, each calculation isolates the effect of one property rather than representing a specific material property set. The reference values, and complete material-property set calculations are documented in Appendix~\ref{app:material_inputs}.

Figure~\ref{fig:kappa_sweep} shows the effect of thermal conductivity. The upper panel compares the focused beamlet radius at the limiting field, $R_f(G_{\mathrm{safe}},B)$, with the thermal diffusion length defined in Eq.~\eqref{diffusion_length}. The lower panel shows the corresponding $G_{\mathrm{safe}}$. These panels show how thermal conductivity impacts the response as magnetic focusing increasingly concentrates the deposited power. At lower magnetic fields, varying $\kappa$ produces a comparatively small change in $G_{\mathrm{safe}}$. In this regime, the larger beamlet distributes the deposited power over a broader area, reducing the importance of heat transport away from the collision centroid. As the magnetic field increases and the beamlet radius decreases, the separation between the curves becomes larger. Increasing $\kappa$ increases the thermal diffusivity through Eq.~\eqref{thermal_diffusivity}, allowing heat to spread farther from the concentrated impact region during the rf pulse. The effect of thermal conductivity on $G_{\mathrm{safe}}$ is therefore stronger after diffusion saturation limits under stronger external magnetic fields.

\begin{figure}[tb!]
\centering
\includegraphics[width=\linewidth]{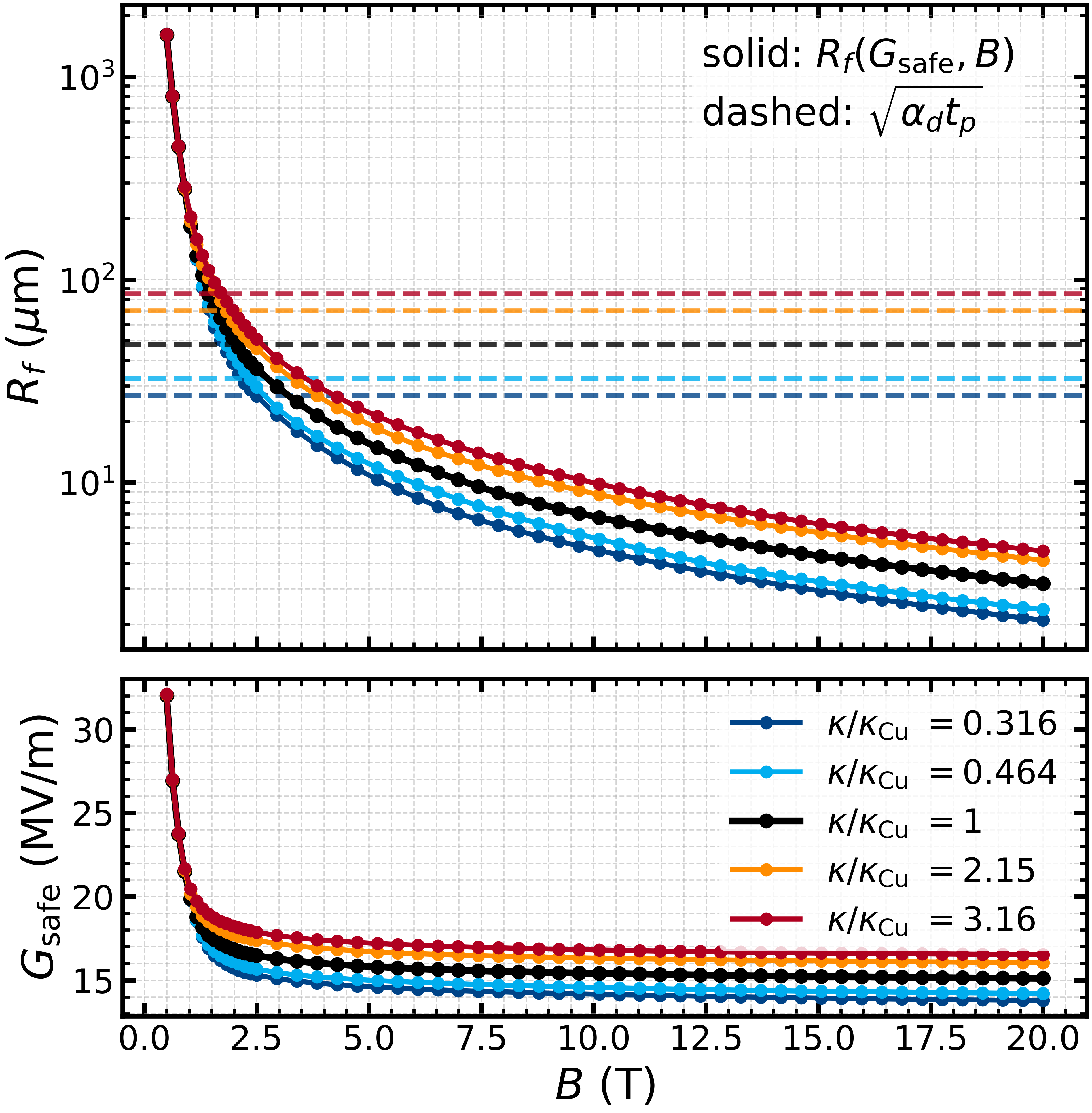}
\caption{
Effect of thermal conductivity relative to the NIST-based Cu reference value at 300~K. The upper panel compares $R_f(G_{\mathrm{safe}},B)$ with the corresponding thermal diffusion length. The lower panel shows $G_{\mathrm{safe}}$ as $\kappa$ is varied with the remaining properties held fixed.
}
\label{fig:kappa_sweep}
\end{figure}

\begin{figure}[tb!]
\centering
\includegraphics[width=\linewidth]{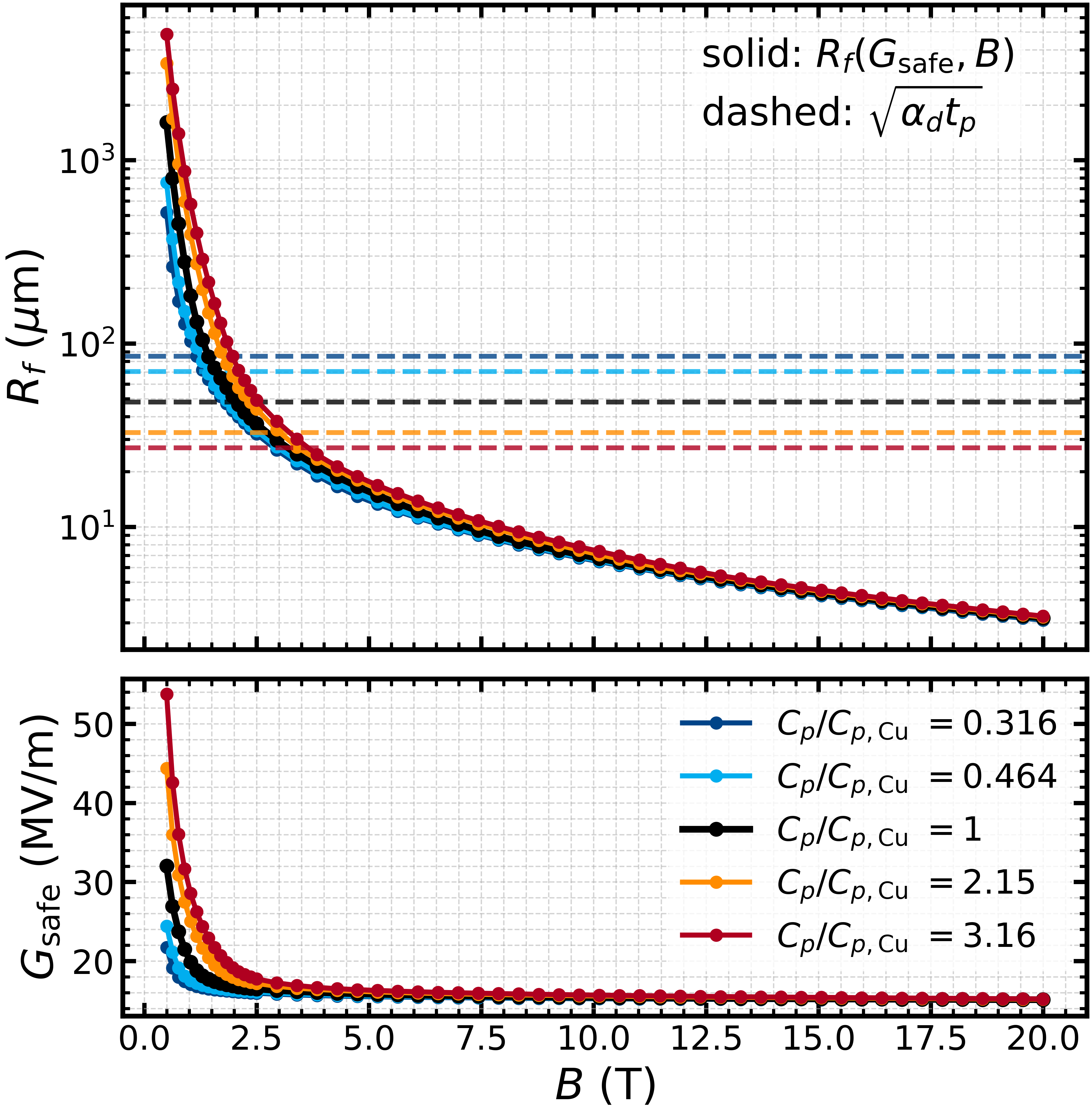}
\caption{
Effect of specific heat capacity relative to the NIST-based Cu reference value at 300~K. The upper panel compares $R_f(G_{\mathrm{safe}},B)$ with the corresponding thermal diffusion length. The lower panel shows $G_{\mathrm{safe}}$ as $C_p$ is varied with the remaining properties held fixed.
}
\label{fig:cp_sweep}
\end{figure}

Figure~\ref{fig:cp_sweep} shows the effect of specific heat capacity. At lower magnetic fields, the deposited power is distributed over a comparatively broad area. Increasing $C_p$ increases the energy required to produce a given temperature rise, improving the material's thermal resilience and raising $G_{\mathrm{safe}}$. However, increasing $C_p$ also reduces the thermal diffusivity when $\kappa$ and $\rho$ are held fixed, so this benefit does not increase uniformly. As the magnetic field increases, magnetic focusing reduces the beamlet radius. Once $R_f(G_{\mathrm{safe}},B)$ becomes smaller than the thermal diffusion length, the concentrated deposition produces a more rapid local temperature buildup, while the reduced diffusivity limits heat redistribution. The $C_p$ curves consequently converge, and varying the specific heat capacity no longer produces an appreciable change in $G_{\mathrm{safe}}$ in the strongly focused regime.

\begin{figure}[tb!]
\centering
\includegraphics[width=\linewidth]{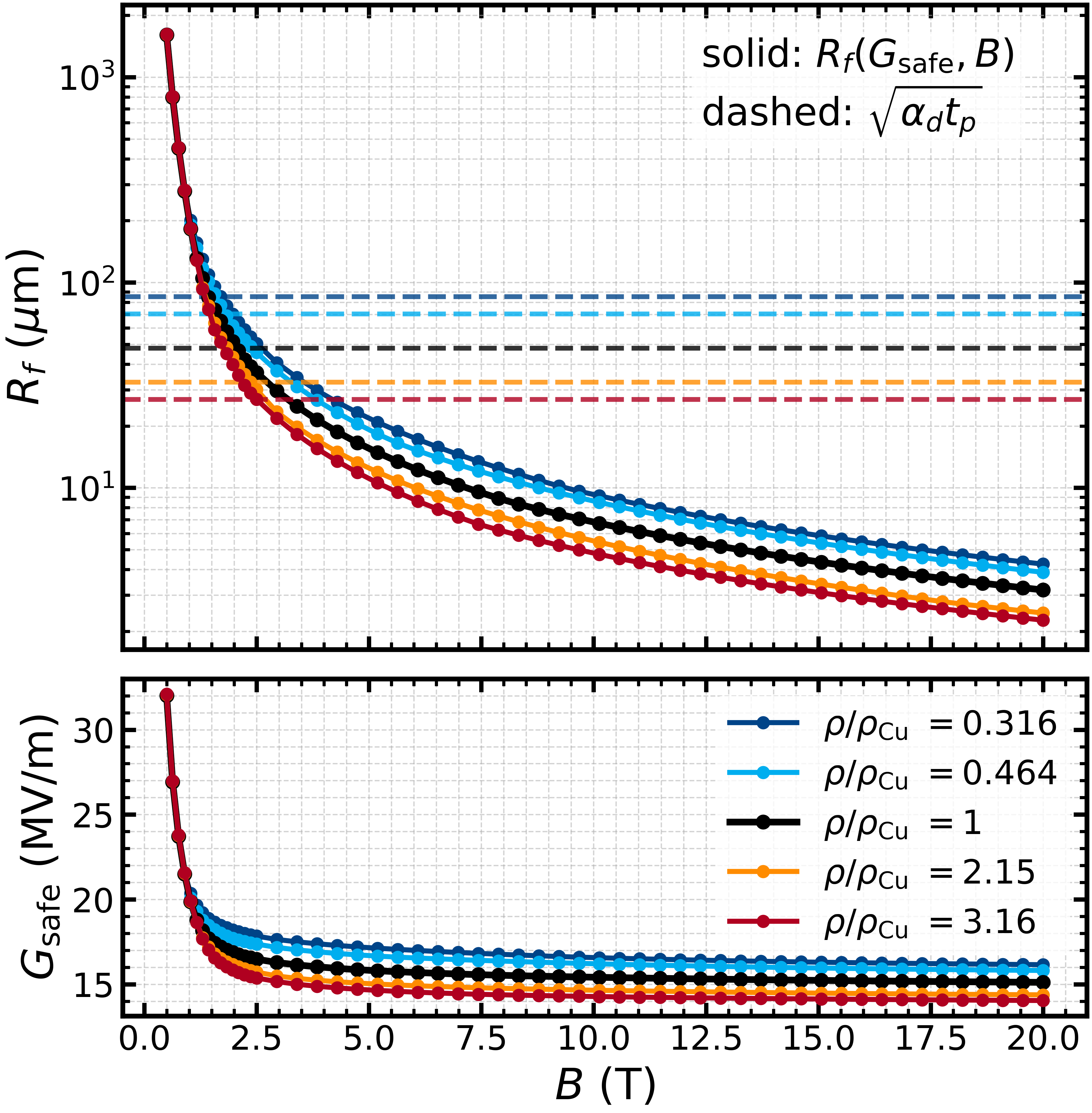}
\caption{
Effect of density relative to the Cu reference value at 300~K. The upper panel compares $R_f(G_{\mathrm{safe}},B)$ with the corresponding thermal diffusion length. The lower panel shows $G_{\mathrm{safe}}$ as $\rho$ is varied in the thermal-response calculation with the deposited-power source held fixed.
}
\label{fig:rho_sweep}
\end{figure}

Figure~\ref{fig:rho_sweep} shows the effect of density. Unlike the $C_p$ and $\alpha_d$ sweeps, varying $\rho$ affects both the electron-bombardment source and the subsequent thermal response. Within the thermal model, density affects both the volumetric heat capacity, $\rho C_p$, therefore the thermal diffusivity in Eq.~\eqref{thermal_diffusivity}, while simultaneously changing the energy deposition and penetration depth in Eqs.~\eqref{eq:bethe_moller} and \eqref{eq:penetration_depth}, respectively. At fixed $\kappa$ and $C_p$, increasing $\rho$ reduces $\alpha_d$ and slows the redistribution of heat away from the collision centroid, promoting localized heat buildup. This effect produces a comparatively small change in $G_{\mathrm{safe}}$ at lower magnetic fields, where the deposited power is distributed over a larger impact area. As the magnetic field increases and the beamlet radius decreases, restricted heat spreading from the concentrated impact region becomes more important, producing greater separation between the density cases. The influence of density is therefore most apparent in the strongly focused regime, where localized heat transport more strongly affects the peak temperature rise.

Together, Figs.~\ref{fig:kappa_sweep}--\ref{fig:rho_sweep} show two regimes governed by the relationship between the focused beamlet radius and the thermal diffusion length. At lower magnetic fields, the deposited power is distributed over a broader area, and increasing $C_p$ raises the energy required for a given temperature rise, thereby increasing $G_{\mathrm{safe}}$. As magnetic focusing drives the beamlet radius below the diffusion length, this specific-heat benefit saturates and the $C_p$ cases converge. Heat transport then becomes increasingly important: higher $\kappa$ promotes heat spreading, whereas higher $\rho$ reduces $\alpha_d$ and restricts redistribution away from the collision centroid. Variations in $\kappa$ and $\rho$ therefore produce their largest changes in $G_{\mathrm{safe}}$ in the strongly focused, higher-field regime. These trends provide the basis for interpreting the combined Cu, Cu 77~K, and Be property sets in the following section.

\subsection*{Breakdown gradient dependence on the rf cavity materials}
\label{sec:nist_material_comparison}

High-gradient experiments motivate the comparison of Cu, cryogenic Cu, and Be. Beryllium cavity elements have supported high-gradient operation in multi-tesla solenoidal fields \cite{bd-bowring,Derun-Be}, while cryogenic-copper experiments have demonstrated improved performance under high electric fields and temperature-dependent field-emission behavior \cite{cuag-cryo,ccfe}. These observations motivate separating changes in the bulk material response from changes in the field-emission source.

\begin{figure}[tb!]
\centering
\includegraphics[width=\linewidth]{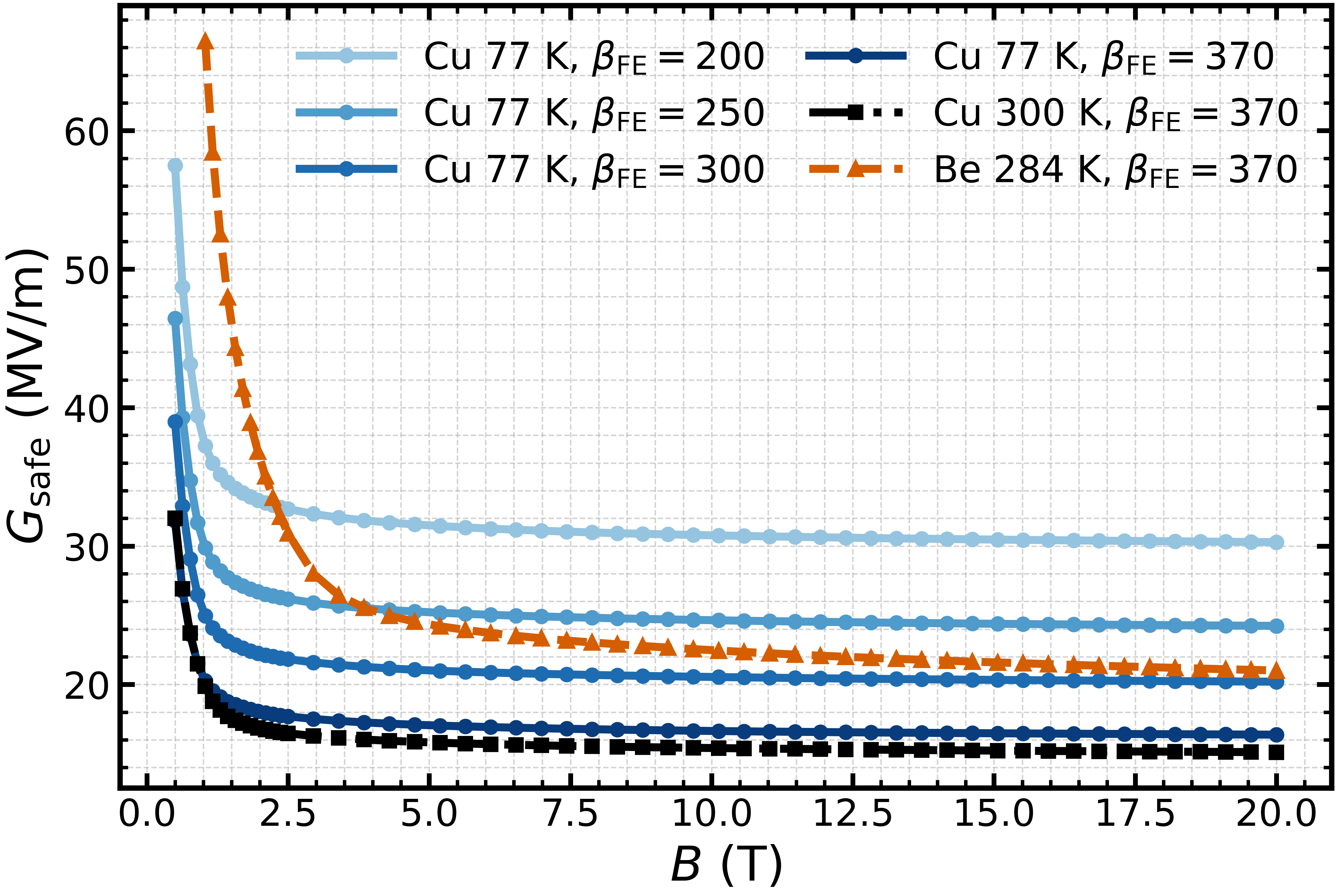}
\caption{
Model-predicted limiting field as a function of magnetic field for the Cu 300~K and Be property sets at $\beta_{\mathrm{FE}}=370$, and for the Cu 77~K property set at the indicated field-enhancement factors. All cases use $\tau_{\mathrm{rf}}=20~\mu\mathrm{s}$. The three $\beta_{\mathrm{FE}}=370$ curves provide the fixed-field-enhancement material comparison, while the additional Cu 77~K curves show the sensitivity to the emission source.
}
\label{fig:nist_material_comparison}
\end{figure}

Figure~\ref{fig:nist_material_comparison} compares the model-predicted limiting field for the Cu 300~K, Cu 77~K, and Be property sets. The baseline material comparison uses $\beta_{\mathrm{FE}}=370$ and $\tau_{\mathrm{rf}}=20~\mu\mathrm{s}$ for all three cases. Additional Cu 77~K curves vary only $\beta_{\mathrm{FE}}$, allowing the sensitivity to the emission source to be evaluated while the cryogenic bulk properties remain fixed. The property values, sources, and calculations are documented in Appendix~\ref{app:material_inputs}.

At the baseline field-enhancement factor, the Be property set produces the highest $G_{\mathrm{safe}}$ across the magnetic-field range considered. The material-specific work function is retained in the Fowler--Nordheim calculation, so the higher assumed Be work function reduces its emitted current relative to Cu. The Be result therefore reflects the combined effects of its emission source, electron energy deposition, heat transport, and thermo-mechanical threshold. Its higher predicted limit should not be attributed to thermal transport or strengthening alone.

The Cu 77~K property set also produces a higher $G_{\mathrm{safe}}$ than the Cu 300~K reference at $\beta_{\mathrm{FE}}=370$, though it is not substantial for the same emitter current. The main benefit is at higher loading concentrations due to Cu 77~K's enhanced diffusivity. The two Cu cases use the same work function, emission area, field-enhancement factor, and yield strength, so their emission currents are identical at a given electric field. Their separation is therefore produced by the temperature-dependent transport, heat-capacity, thermal-expansion, and elastic inputs used in the model. These properties are evaluated at the initial material temperature and held fixed during the rf pulse; the calculation does not include their subsequent variation as the impact region heats.

Experimental evidence for temperature-dependent field emission motivates the additional Cu 77~K field-enhancement cases \cite{ccfe}. Because the Fowler--Nordheim current depends strongly on $\beta_{\mathrm{FE}}$, reducing the effective field-enhancement factor substantially decreases the electron-bombardment source. At the common reference field $E_0=30~\mathrm{MV/m}$, the rf-cycle-averaged Cu emission current decreases from $27.554~\mathrm{mA}$ at $\beta_{\mathrm{FE}}=370$ to $4.400$, $0.698$, and $0.0500~\mathrm{mA}$ at $\beta_{\mathrm{FE}}=300$, 250, and 200, respectively. These values illustrate the source sensitivity at a fixed field; in each $G_{\mathrm{safe}}$ calculation, the current is recalculated self-consistently at the applied field.

The additional Cu 77~K curves show that reducing $\beta_{\mathrm{FE}}$ raises the predicted limiting field beyond the improvement produced by the cryogenic bulk properties alone. This is a conditional sensitivity test: it does not assume that cooling produces a particular reduction in $\beta_{\mathrm{FE}}$ or identify the surface mechanism responsible. Instead, the comparison separates the modeled consequence of a weaker emission source from the fixed-property Cu 77~K result. Figure~\ref{fig:nist_material_comparison} therefore distinguishes the improvement that would result if cryogenic operation or conditioning also reduced the effective field enhancement.

\subsection*{Breakdown gradient dependence on the rf pulse length}
\label{sec:pulse_length_sensitivity}

\begin{figure}[tb!]
\centering
\includegraphics[width=\linewidth]{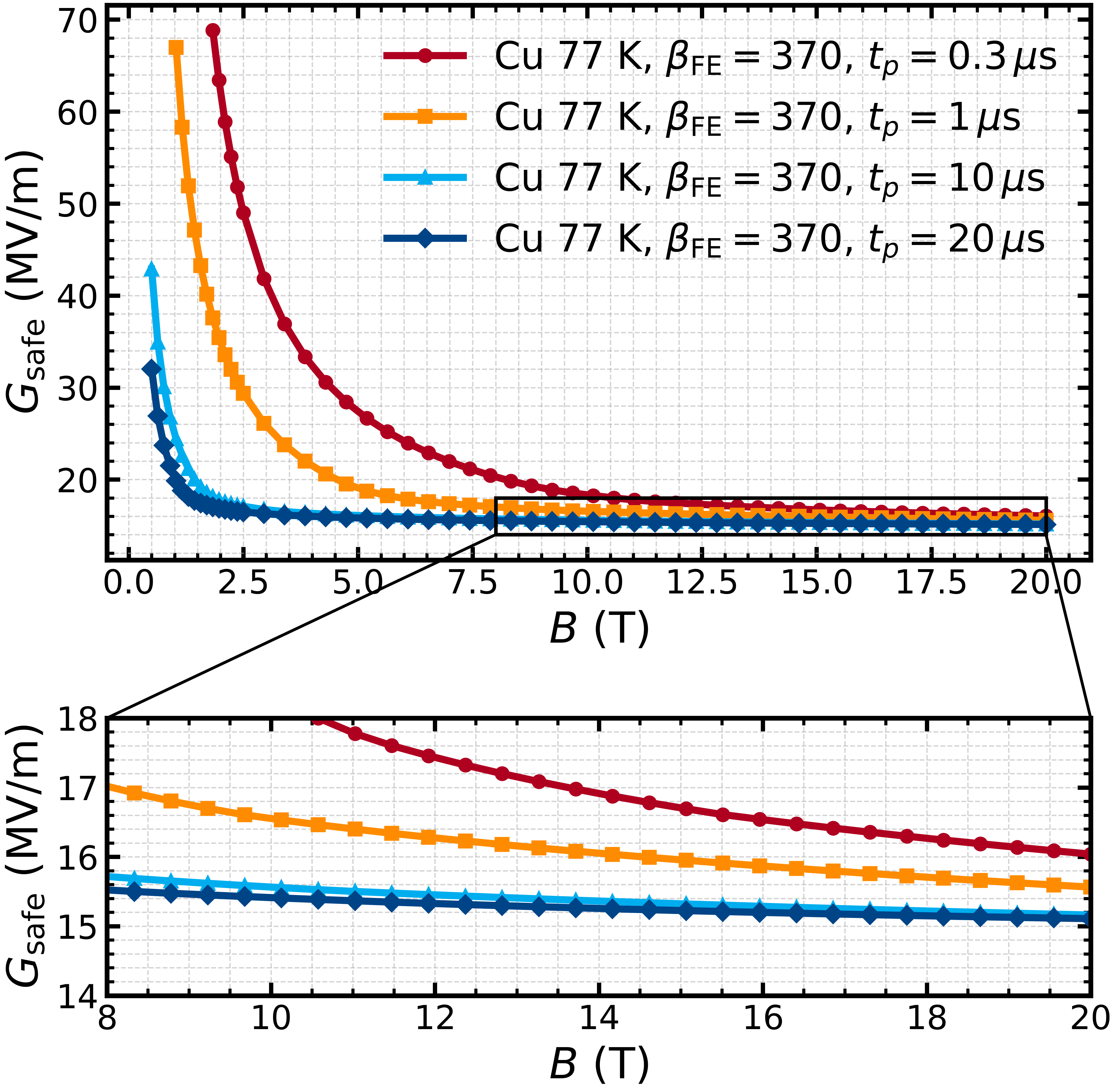}
\caption{
Effect of rf pulse length on the model-predicted limiting field for NIST-based Cu at 77~K with $\beta_{\mathrm{FE}}=370$. The pulse length is varied while the remaining model inputs are held fixed. The lower panel shows a zoomed-in view of the high external mangetic field region from 8-20~T, showing the dwindling benefit of pulse length reduction as the external magnetic field increases.
}
\label{fig:pulse_length_sog}
\end{figure}
Recent short-pulse high-gradient studies motivate examining the dependence on rf pulse length. Prior work has reported empirical scaling of breakdown with accelerating field and pulse duration~\cite{grudiev-short-pulse}, demonstrated very high surface electric fields using nanosecond rf pulses~\cite{tan-short-pulse}, and examined dark-current behavior in short-pulse high-gradient structures~\cite{rijal-short-pulse}.  Although those studies concern breakdown regimes and mechanisms distinct from the magnetically focused electron bombardment considered here, they motivate assessing how the thermal response predicted by this model depends on pulse length.

Figure~\ref{fig:pulse_length_sog} shows the model-predicted limiting field for the NIST-based Cu property set at 77~K with $\beta_{\mathrm{FE}}=370$ and pulse lengths of $0.3$, $1$, $10$, and $20~\mu\mathrm{s}$. All other model inputs are held fixed. Shortening the pulse reduces the duration over which the field-emitted electrons deposit energy and also reduces the thermal diffusion length through Eq.~\eqref{diffusion_length}. Reducing the pulse length raises $G_{\mathrm{safe}}$ over part of the magnetic-field range. At higher magnetic fields, however, the curves approach a common limiting trend as the beamlet becomes strongly focused. Within the electron-bombardment mechanism modeled here, the benefit of reducing the exposure time therefore saturates in this regime. Consistent with the length-scale behavior shown in Figs.~\ref{fig:kappa_sweep}--\ref{fig:rho_sweep}, shorter pulses can raise the predicted limiting field but cannot indefinitely compensate for the increasing concentration of deposited power at higher magnetic fields.

\section{Discussion on caveats and limitations}
\label{sec:discussion}

The results presented here are model-dependent estimates of the limiting field associated with magnetically focused electron bombardment and should therefore not be interpreted as absolute rf breakdown limits or as a complete description of breakdown in an operating cavity.

The tracking calculation is restricted to electrons emitted from a selected high electric field source and followed until collision with a cavity surface. It does not include secondary cascades, repeated emission from impact sites, transmission through the beryllium windows, or transport between multiple cavities. The calculated collision distributions therefore apply to the selected first-cavity configuration and do not represent particle transport throughout the complete cooling cell. The emission source is also held fixed during tracking. Changes in the emitter geometry, field-enhancement factor, emitted current, or number of active emission sites during operation are not modeled. Possible cavity loading by larger dark currents is likewise not evaluated in this work.

The energy-deposition treatment introduces additional approximations. The stopping model in Eq.~\eqref{eq:bethe_moller} accounts for collisional energy loss but does not include radiative losses such as bremsstrahlung. The penetration depth is also prescribed using Eq.~\eqref{eq:penetration_depth} rather than obtained from a full particle-transport calculation. These effects can become increasingly relevant as the incident electron energy increases. A more complete treatment of the deposited-energy distribution would require coupled electromagnetic particle transport including both collisional and radiative interactions. The NIST database from Ref. \cite{nist-data} indicates that the radiative effects become more significant at higher energies.
However, for the incident-electron energies reached at the limiting fields predicted here, radiative losses are expected to remain subdominant.

The temperature-rise calculation holds the material properties fixed throughout each rf pulse and consequently does not include temperature-dependent changes in thermal transport, elastic response, thermal expansion, or yield strength. Transient changes in surface morphology are also excluded. In addition, the copper yield strength is treated as a fixed reference value, although it can depend on material treatment, impurity content, and conditioning history.

\section{Conclusion}
\label{sec:conclusion}

This work developed a tracking and thermal-response framework for examining magnetically focused field-emitted electrons in rf cavities operating within strong external magnetic fields. The two-phase tracking approach resolves the enhanced fields and space-charge-driven expansion near the emission site before transporting the resulting electron distribution through the cavity-scale electromagnetic fields. In the reduced-field configuration, the simulated beamlet radius agrees well with the analytical theory of the simplified model of Ref.~\cite{bd-diktys}. When the full rf eigenmode fields and spatially varying external magnetic field are included, the collision distributions exhibit phase-dependent centroid motion and deformation of the approximately circular beamlet structure. Nevertheless, the impacts remain localized relative to the cavity dimensions, supporting the use of the reduced analytical model as a characteristic description of the localized deposited-power source.

Using this analytical model, the thermal-response study examined how the predicted thermomechanical operating gradient limit depends on the thermal and mechanical properties entering the model. Variations in thermal conductivity, specific heat capacity, and density demonstrate that the influence of individual material properties changes as magnetic focusing reduces the characteristic impact radius relative to the thermal diffusion length. The calculations therefore show that the response cannot generally be attributed to a single thermal or mechanical property, but instead depends on the coupled effects of localized energy deposition, heat storage, thermal transport, and the thermomechanical threshold. The model was subsequently applied to representative Cu, cryogenic Cu, and Be property sets and to different rf pulse lengths to illustrate how material selection and operating conditions affect the predicted limiting field. These comparisons provide guidance for evaluating material and operating-condition choices for future muon-cooling cavities in strong magnetic fields and other rf systems (e.g. photocathodes) operated in external magnetic fields. The calculated gradient limits should be interpreted as comparative predictions for the magnetically focused electron-bombardment mechanism rather than as absolute rf breakdown thresholds. Further predictive application will require extensions that include temperature-dependent material behavior, evolution of the emitting surface, and additional particle-transport processes.

\begin{acknowledgments}
The authors thank Dr. Gaoxue Wang for providing the \textit{ab initio} data used in the DFT alloy calculations, and Drs. Alexej Grudiev, Carmelo Barbagallo, Chris Rogers, and Ruihu Zhu of the IMCC for providing the realistic field maps, and Dr. Ji Qiang, Dr. Omkar Ramachandran, Dr. Michael Shapiro, Dr. Chengkun Huang, Thomas Harless, and Gaurab Rijal for valuable discussions.
This research was supported by the U.S. Department
of Energy, Office of Science, Office of High Energy
Physics under Awards No. DE-SC0021928, No. DE-SC0014664, and DOE Contract No.~DE-AC02-05CH11231.

\end{acknowledgments}

\appendix

\section{Field-Emission Source and Macroparticle Allocation}
\label{app:field_enhancement}

This appendix summarizes the numerical allocation procedure used to convert the field-emission source into a time-dependent macroparticle distribution for rf~Track \cite{latina_rf_track}.
The main tracking section describes the physical setup. Here, the purpose is to document how the enhanced local electric field is converted into emitted charge and how that charge is sampled into particles.

The emitting feature is modeled as a prolate spheroidal asperity with semi-major axis $a=60~\mu\mathrm{m}$ and semi-minor axis $b=1.77~\mu\mathrm{m}$. The cavity rf electric field at the emission site is obtained from the rf eigenmode map. The analytical asperity electric-field contribution is added to the imported cavity rf electric field to define the local emission electric field,
$\mathbf{E}{\mathrm{loc}}=\mathbf{E}{\mathrm{rf}}+\mathbf{E}{\mathrm{asp}}$,
where $\mathbf{E}{\mathrm{asp}}$ is calculated from the prolate-spheroid field-enhancement model used in Ref.~\cite{bd-diktys}. The field-enhancement factor is defined as $\beta_{\mathrm{FE}}=|\mathbf{E}{\mathrm{loc}}|/|\mathbf{E}{\mathrm{rf}}|$. For the asperity geometry used in this work, the maximum field enhancement is approximately $\beta_{\mathrm{FE}}=357$.

The local emission electric field is evaluated over the rf cycle at the Stage~B8 frequency, $\omega_{\mathrm{rf}}=2\pi f_{\mathrm{rf}}$, such that the instantaneous field is $E_{\mathrm{em}}(t)=\operatorname{Re}\left[E_{\mathrm{loc}}e^{i\omega_{\mathrm{rf}}t}\right]$. Emission is allowed only during the accelerating phase, when the local electric field points outward from the emitting surface. The instantaneous emission current density is then calculated using the Fowler--Nordheim relation in Eq.~\eqref{fn_eq} with the instantaneous local emission field $E_{\mathrm{em}}(t)$.

The emitting surface is discretized into cells with area $A_j$. For a time bin $t_i$ with width $\Delta t$, the charge emitted from surface cell $j$ is $\Delta Q_{ij}=J_{\mathrm{FN}}(E_{\mathrm{em},ij})A_j\Delta t$. The total charge emitted in time bin $i$ is $Q_i=\sum_j\Delta Q_{ij}$, and the total emitted charge over the rf emission window is $Q_{\mathrm{tot}}=\sum_i Q_i$.

A fixed total number of macroparticles, $N_{\mathrm{macro}}$, is generated from this emitted-charge distribution. The probability that a macroparticle is assigned to time bin $i$ is $P_i=Q_i/Q_{\mathrm{tot}}$, so time bins with larger emitted charge receive more macroparticles. After a macroparticle is assigned to time bin $i$, its emitting surface cell is selected using the conditional probability $P_{j|i}=\Delta Q_{ij}/Q_i$. This assigns more macroparticles to regions of the asperity surface where the Fowler--Nordheim current density is larger. Once a cell is selected, the macroparticle position is randomized within that surface element to avoid initializing multiple particles at identical coordinates. The emission time is assigned from the selected rf time bin.

\begin{figure}[tb!]
\centering
\includegraphics[width=\linewidth]{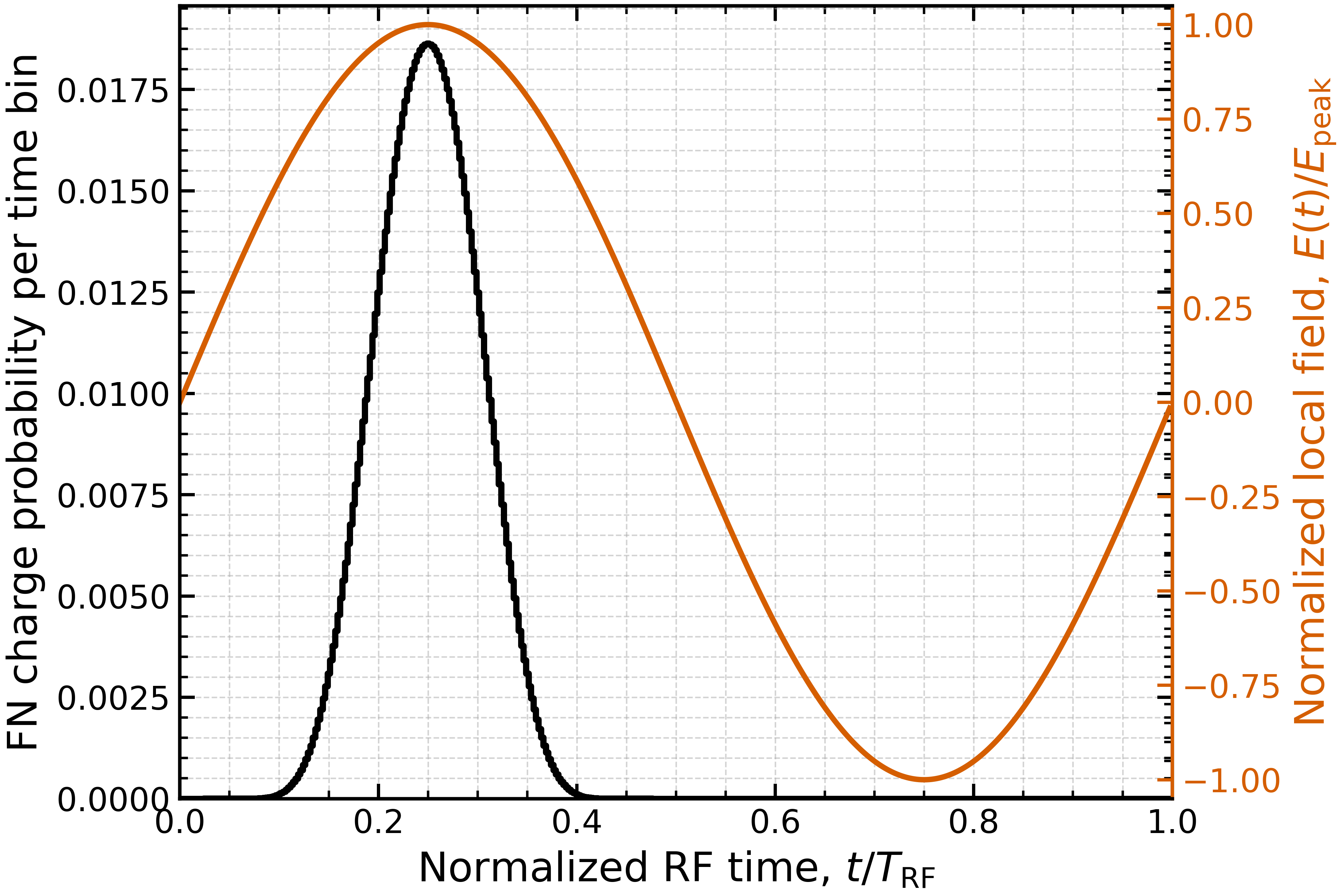}
\caption{
Normalized instantaneous local emission electric field and corresponding Fowler--Nordheim charge-sampling probability over one rf cycle. The horizontal coordinate is normalized to the rf period, $t/T_{\mathrm{rf}}$.
}
\label{fig:fn-sampling}
\end{figure}

Figure~\ref{fig:fn-sampling} illustrates the temporal component of the emission sampling. Because the Fowler--Nordheim current depends exponentially on the instantaneous local emission electric field, most of the emitted charge is concentrated near the peak of the outward-accelerating portion of the rf cycle.

All macroparticles are assigned the same charge weight, $q_{\mathrm{macro}}=Q_{\mathrm{tot}}/N_{\mathrm{macro}}$, corresponding to $N_e^{(\mathrm{macro})}=q_{\mathrm{macro}}/e$ physical electrons per macroparticle. Thus, the spatial and temporal structure of the emitted current is represented by the macroparticle sampling density, while each macroparticle carries the same physical charge.

Each emitted particle is initialized with a kinetic energy $K_0=1$~eV directed along the local emission direction. The corresponding launch momentum is $p_0c=m_ec^2\sqrt{(1+K_0/m_ec^2)^2-1}$. The resulting time-dependent bunch is then passed to rf~Track. Space charge is enabled during the initial enhanced-field tracking stage, so the emitted distribution evolves under the imposed tracking fields and its self-fields before being passed to the full cavity tracking calculation.

\section{Material-property inputs}
\label{app:material_inputs}
\begin{table*}[tb!]
\caption{
Relative thermal and thermo-mechanical inputs used in the model.
The NIST/reference Cu and Be cases are used in the material comparison, while the DFT-based pure-Cu and CuAg cases are included as reference property sets.
All values are normalized to the NIST-based Cu reference at 300~K. DFT-based values were provided by the authors of Refs.~\cite{alloy1,alloy2}.
}
\label{tab:relative_material_inputs}
\centering
\footnotesize
\begin{ruledtabular}
\begin{tabular*}{\textwidth}{@{\extracolsep{\fill}}lclcccccc}
Case
& $T$ (K)
& Source
& $\rho/\rho_{\rm Cu,NIST}^{300K}$
& $\kappa/\kappa_{\rm Cu,NIST}^{300K}$
& $C_p/C_{p,\rm Cu,NIST}^{300K}$
& $\alpha_d/\alpha_{d,\rm Cu,NIST}^{300K}$
& $\alpha_{\rm th}/\alpha_{\rm th,Cu,NIST}^{300K}$
& $\Delta T_s/\Delta T_{s,\rm Cu,NIST}^{300K}$ \\
\hline
Cu
& 300
& NIST
& 1.000
& 1.000
& 1.000
& 1.000
& 1.000
& 1.000 \\

Cu
& 77
& NIST
& 1.000
& 1.443
& 0.503
& 2.869
& 0.479
& 1.944 \\

Be
& 284
& NIST
& 0.206
& 0.680
& 4.678
& 0.705
& 0.639
& 3.626 \\

Cu$_{\rm dft}$
& 300
& DFT
& 0.982
& 0.799
& 0.995
& 0.818
& 1.118
& 1.219 \\

Cu$_{\rm dft}$
& 77
& DFT
& 0.982
& 1.519
& 0.519
& 2.983
& 0.541
& 2.352 \\

CuAg$_{4,\rm dft}$
& 300
& DFT
& 0.999
& 0.752
& 0.940
& 0.800
& 1.471
& 3.344 \\

CuAg$_{4,\rm dft}$
& 77
& DFT
& 0.999
& 1.061
& 0.529
& 2.009
& 0.682
& 6.486 \\
\end{tabular*}
\end{ruledtabular}
\end{table*}

The NIST-based reference cases use the published correlations for OFHC Cu and polycrystalline Be \cite{nist-data}. For properties represented by the NIST logarithmic polynomial, the value at temperature $T$ is calculated as

\begin{equation}
\log_{10}y=\sum_{n=0}^{8}a_n\left(\log_{10}T\right)^n.
\label{eq:nist_log_polynomial}
\end{equation}

The coefficients and units are specific to each property. This form is used for the Cu specific heat capacity and thermal expansion coefficient and for the Be specific heat capacity.

The thermal conductivity of OFHC Cu is calculated from the NIST correlation

\begin{equation}
\log_{10}\kappa_{\mathrm{Cu}}=
\frac{a+cT^{1/2}+eT+gT^{3/2}+iT^2}
{1+bT^{1/2}+dT+fT^{3/2}+hT^2},
\label{eq:nist_cu_thermal_conductivity}
\end{equation}

using the coefficients for $\mathrm{RRR}=500$.

For Be, NIST provides the polycrystalline relative linear-expansion correlation \cite{nist-data},
\begin{equation}
y_{\mathrm{Be}}(T)=a+bT+cT^2+dT^3+eT^4,
\end{equation}
where
\begin{equation}
y_{\mathrm{Be}}(T)=10^5\frac{L(T)-L_{293}}{L_{293}}.
\end{equation}
Applying the definition $\alpha_{\mathrm{th}}=L^{-1}dL/dT$ and using a small-strain approximation $L(T)\approx L_{293}$ gives the form implemented in the calculation,
\begin{equation}
\alpha_{\mathrm{th,Be}}(T)
\approx
10^{-5}\left(b+2cT+3dT^2+4eT^3\right).
\label{eq:nist_be_thermal_expansion}
\end{equation}

The Cu properties are evaluated at 300 and 77~K. The NIST correlation used for the Be specific heat capacity is defined over 14--284~K. The Be reference case is therefore evaluated at 284~K rather than extrapolating the correlation to 300~K. The fixed densities used in the calculation are $\rho_{\mathrm{Cu}}=8.96$~g/cm$^3$ and $\rho_{\mathrm{Be}}=1.848$~g/cm$^3$, and the Be reference case uses $\kappa_{\mathrm{Be}}=2.73$~W/(cm,K). For each case, the thermal diffusivity is calculated using Eq.~\eqref{thermal_diffusivity}, and the thermo-mechanical threshold is calculated using Eq.~\eqref{tthresh}. The yield strength is fixed at $\sigma_y=62$~MPa for both Cu cases and at $\sigma_y=240$~MPa for the Be case.

The density functional theory (DFT) based pure-Cu and CuAg inputs were provided by the authors of Refs.~\cite{alloy1,alloy2}. These values are predictions for the compositions, temperatures, and computational conditions considered in those studies. They are not experimental measurements or exact material constants.

For a copper alloy $\mathrm{Cu_{N_{tot}-N_X}X_{N_X}}$, the solute concentration is defined as

\begin{equation}
c=\frac{N_{\mathrm{X}}}{N_{\mathrm{tot}}},
\label{eq:dft_composition}
\end{equation}

where $N_{\mathrm{X}}$ and $N_{\mathrm{tot}}$ are the numbers of solute and total atoms in the DFT supercell, respectively. Composition-weighted atomic inputs are calculated as

\begin{equation}
P_{\mathrm{CuX}}(c)=(1-c)P_{\mathrm{Cu}}+cP_{\mathrm{X}},
\label{eq:dft_weighted_property}
\end{equation}

where $P$ represents the work function, atomic mass, atomic number, or mean excitation energy used in the field-emission and energy-deposition calculations.

The density is calculated from the composition-weighted atomic mass and DFT supercell volume,

\begin{equation}
\rho_{\mathrm{CuX}}=
\frac{N_{\mathrm{tot}}A_{\mathrm{at,CuX}}m_u}{V_{\mathrm{cell}}},
\label{eq:dft_density}
\end{equation}

where $m_u$ is the atomic mass unit and $V_{\mathrm{cell}}$ is the supercell volume.

The thermal conductivity is calculated from the supplied electrical conductivity using the Wiedemann--Franz relation,

\begin{equation}
\kappa_{\mathrm{CuX}}(c,T)=L_0\sigma_{\mathrm{el,CuX}}(c,T)T,
\label{eq:dft_thermal_conductivity}
\end{equation}

where $L_0$ is the Lorenz number.

The Young's modulus is calculated from the DFT-based bulk and shear moduli according to

\begin{equation}
E_{\mathrm{mod}}=\frac{9KG}{3K+G},
\label{eq:dft_youngs_modulus}
\end{equation}

and Poisson's ratio is calculated as

\begin{equation}
\nu=\frac{3K-2G}{2(3K+G)}.
\label{eq:dft_poisson_ratio}
\end{equation}

For the CuAg case, the predicted critical-resolved-shear-stress increment, $\tau$, is converted to a yield-strength increment using

\begin{equation}
\Delta\sigma=M\tau,
\label{eq:dft_yield_increment}
\end{equation}

where $M=3.06$ is the Taylor factor used for the FCC lattice. The resulting alloy yield strength is

\begin{equation}
\sigma_y(c,T)=\sigma_{y,\mathrm{Cu}}+\Delta\sigma(c,T).
\label{eq:dft_yield_strength}
\end{equation}

The resulting density, thermal conductivity, specific heat capacity, and thermal diffusivity are applied in the temperature-rise calculation in Eq.~\eqref{trise}. The elastic modulus, Poisson ratio, thermal expansion coefficient, and yield strength determine the threshold in Eq.~\eqref{tthresh}. The DFT-based thermo-mechanical properties for CuAg are retained as reference property sets in Table \ref{tab:relative_material_inputs} and are not included in the NIST-based material comparison in Fig.~\ref{fig:nist_material_comparison}.

\bibliographystyle{apsrev4-2}
\bibliography{bib}
\end{document}